\documentstyle[twoside,leqno,epsf,11pt]{article}  

\newcommand{\heading}[1]{{\vspace{8pt}\noindent\sc{#1}}}

 \newtheorem{definition}{Definition}

 \newtheorem{corollary}{Corollary}
 \newtheorem{theorem}{Theorem}
 \newtheorem{lemma}{Lemma}
 \newtheorem{fact}{Fact}

\newenvironment{proof}{\par \sf Proof.\rm}{\hspace*{\fill}$\Box$\vspace{1ex}}
 \newcommand{\ignore}[1]{}

\newcommand{\half}{\frac{1}{2}}

\newcommand{\agent }{agent }
\newcommand{\agents }{agents }
\begin{document}

\def\ourtitle{{\bf Mutual Search\thanks{A preliminary version
of this work was presented at the 
{\em 9th ACM-SIAM
     Symposium on Discrete Algorithms}, San Francisco, January 1998}}
 }

\title{{\LARGE \ourtitle}\\
\protect\vspace{0.15in}}

\author{Harry Buhrman\thanks{
Partially supported by  NWO through NFI Project
ALADDIN under contract number NF 62-376 
and by the European Union
through NeuroCOLT ESPRIT Working Group Nr. 8556.
Centrum voor Wiskunde en Informatica (CWI),
Kruislaan 413, 1098 SJ Amsterdam, The Netherlands.
{\tt buhrman@\allowbreak cwi.nl}.}
\and Matthew Franklin\thanks{
   Xerox PARC,
   3333 Coyote Hill Road,
   Palo Alto, CA 94304, USA.
{\tt franklin@\allowbreak parc.xerox.com.}}
\and Juan A. Garay\thanks{
Bell Labs - Lucent Technologies,
   600 Mountain Ave, 
   Murray Hill, NJ 07974,
   USA.
{\tt garay@\allowbreak research.bell-labs.com}. 
Partially supported by  NWO through NFI Project
ALADDIN under contract number NF 62-376 and by a SION grant
and by the European Union
through NeuroCOLT ESPRIT Working Group Nr. 8556 while the
author was visiting CWI.}
\and Jaap-Henk Hoepman\thanks{University Twente,
Dept. of Computer Science, Box 217, 7500 AE Enschede, The Netherlands.
{\tt jhh@\allowbreak xs4all.nl}.}
\and John Tromp\thanks{
Partially supported by  NWO through NFI Project
ALADDIN under contract number NF 62-376 
and by the European Union
through NeuroCOLT ESPRIT Working Group Nr. 8556.
Centrum voor Wiskunde en Informatica (CWI),
Kruislaan 413, 1098 SJ Amsterdam, The Netherlands.
{\tt tromp@cwi.nl}. }
\and Paul Vit\'{a}nyi\thanks{
Partially supported by  NWO through NFI Project
ALADDIN under contract number NF 62-376 
and by the European Union
through NeuroCOLT ESPRIT Working Group Nr. 8556.
Centrum voor Wiskunde en Informatica (CWI),
Kruislaan 413, 1098 SJ Amsterdam, The Netherlands.
{\tt paulv@ \allowbreak cwi.nl.}}
}

\date{}

\maketitle

\vspace{.3in}

\def\ourabstract{
We introduce a search problem called ``mutual search''
where $k$ \agents, arbitrarily distributed over $n$ sites, are required to
locate one another by posing 
queries of the form ``Anybody at site $i$?''.
We ask for the least number of queries that is necessary and sufficient.
For the case of two \agents using deterministic
protocols we obtain the following worst-case results:
In an oblivious setting
(where all pre-planned queries are executed) there is no savings:
$n-1$ queries are required and are sufficient. In a nonoblivious setting
we can exploit the paradigm of ``no news is also news''
to obtain significant savings: in the synchronous case
$0.586n$ queries suffice and 
$0.536n$ queries are required;
in the asynchronous case
$0.896n$ queries suffice and a fortiori $0.536$ queries are required;
for $o(\sqrt{n})$ \agents using a deterministic
protocol less than $n$ queries suffice; there is a simple
randomized protocol for two \agents with worst-case expected 
$0.5n$ queries and all randomized protocols
require at least $0.125n$ worst-case expected queries.
The graph-theoretic framework we formulate for expressing and analyzing
algorithms for this problem may be of independent interest.
}

\begin{abstract}
\ourabstract
\end{abstract}

\vspace{.3in}

\section{Introduction}\label{sect.intro}

Search problems 
come in many forms \cite{Kn98}. Perhaps the following one is new:
Suppose you
and a friend 
check separately into the same hotel in Las Vegas in different
rooms. 
For reasons we don't go into here
you both don't want to draw any attention to 
your relation. You are supposed to phone one another at noon, but
unfortunately you don't know each others' room number.
What to do? 
Every room contains a room phone and room number.
You can phone all other rooms in the
hotel to find your friend and she can do the same (if
the wrong person picks up the phone you simply hang up and nobody
is the wiser). 
This will cost a lot of time:
there are 1000 rooms. In the worst case you use almost 2000 
room calls together. Luckily you and your friend know
the protocol in this paper: you locate one another
using only 586 room calls together in the worst case. 
There are more serious problems
of the same nature that are listed
in the Appendix~\ref{app.related}.

In general we can think of $k \geq 2$ \agents having to find
each others' locations in a uniform unstructured search space
consisting of $n$ sites ($n \geq k$).
The sites have
distinct identities, say $0, \ldots , n-1$ ($k \leq n$),
every site can contain zero or one \agent, and the \agents
execute identical protocols based on the values $n,k$ with
their site identity as input. The \agents can execute queries of
the form ``Anybody at site $i$'' and every such query results in
an answer ``yes'' or ``no.''
We say that two \agents know each others'
location as soon as one agent queries the location
of the other agent or the other way around. Before that happens
they don't know each other's location.
The relation ``know location'' is transitive and the problem is solved if
all $k$ \agents know one another's location.
This type of search can be called {\em Mutual Search} ({\em MS}).
We  analyze the cost in number of queries
for the case $k=2$
under various timing assumptions for deterministic
and randomized algorithms. We also give a result for the
general case of $k = o( \sqrt{n} )$ \agents. 

{\bf Our Results}: We first look at {\em deterministic protocols} for
two \agents.  If the protocol is {\em oblivious}, 
so that the cost for each \agent
is a fixed number of queries,
then there are no significant savings possible:  two \agents
need to place at least $n-1$ queries in total in the worst case.
\footnote{``Oblivious'' means that the queries scheduled at certain
time slots take place independent
of the replies received. The same lower bound holds if there is no
FIFO discipline: the answer to a query can arrive only after 
the following queries are executed. This is the case
when the sites are nodes in a computer network,
the \agents are processes at those nodes that query
by sending messages over communication links
with unknown bounded (or possibly unbounded as in the FLP model \cite{FLP})
communication delay without waiting for the answers to earlier messages. Then, 
a process may have to send all its messages before
an affirmative reply to one of the early messages is received.}
Namely, given a protocol, construct the directed graph on $\{0,
\ldots , n-1\}$ with an arc from $i$ to $j$ if an \agent at $i$ queries node
$j$. For every pair there must be at least one arc. Hence there are 
at least ${n
\choose 2}$ arcs in total, and the average number of outgoing 
arcs per node is at
least $\frac{n-1}{2}$. It follows that some pair of nodes must
together have twice this number, or $n-1$, of outgoing edges (this
can be refined to $2\lceil\frac{n-1}{2} \rceil$ for all $n>2$).
The tightness of this bound is witnessed by an algorithm called HalfInTurn,
to be discussed in Section~\ref{simplealgs}:

\noindent
{\bf Oblivious case ($k=2$):} Both upper bound and 
lower bound are $2 \lceil \frac{n-1}{2} \rceil$
queries in the deterministic worst-case.

In the remainder of the paper we analyze the
nonoblivious case. We obtain savings by exploiting the information
inherent in timing (``no news is also news'') and a prescribed
order of events.

\noindent
{\bf Synchronous case ($k=2$):} 
in Section~\ref{smooth} we present the protocol
 SR$_n$, an algorithm with a worst-case cost
of only $(2-\sqrt{2})n \approx 0.586 n$.
We also show this algorithm to be 
close to optimal,
by proving a $(4-2\sqrt{3})n \approx 0.536 n$ lower
bound on the number of queries required by any 
mutual search algorithm in Section~\ref{lowerbound}.

\noindent
{\bf Asynchronous case ($k=2$):} In Section~\ref{asynch}
we show that
there is a mutual search algorithm 
that uses asymptotically $0.896n$ queries. The best lower bound 
we know of is the $0.536 n$ lower bound in Section~\ref{lowerbound}.

\noindent
{\bf Randomized case ($k=2$):} 
We consider {\em randomized algorithms} for the problem in
Section~\ref{random}.
A synchronous randomized algorithm is shown to have a
worst-case (over \agent location) expected (over random coin flips)
cost of about $\frac{n+1}{2}$, thus beating the
deterministic lower bound.
We show a lower bound on the worst-case expected number of
queries of $\frac{n-1}{8}$.

%



\noindent
{\bf Synchronous multi-agent case ($k=o(\sqrt{n}))$:}
In Section~\ref{multiplayer} we present RS$_{n,k}$, 
a deterministic algorithm for $k \geq 2$ \agents
with a cost well below $n$ for all $k=o(\sqrt{n})$.

The framework we develop for reasoning about the Mutual Search
problem may be of independent interest. 
Mutual search can serve as a preliminary stage to sharing random 
resources in a distributed setting or forming
coalitions for Byzantine attacks and various cryptographic settings. \\

\noindent
{\bf Related Work}: The authors believe that this is a novel type of
search problem that has not been considered before.  We do not know of
any directly related previous research.  Several topics that are more
or less related can be found in the 
Appendix~\ref{app.related}.

\section{Synchronous Case For Two Agents}

In Section~\ref{model-sec} we formulate the model for the synchronous
case with $k=2$ agents  located at $n$ sites and give a framework for 
expressing and analyzing the structure
of algorithms for this problem. We analyze this case fairly completely,
but in later sections we also present results for other 
instances of the MS problem.

\subsection{Model and Definitions}
\label{model-sec}

Consider $n$ sites numbered $0, \ldots , n-1$ with $k\leq n$
\agents distributed over the $n$ sites with zero or one \agent
per site.
Time is discrete, with time slots numbered $0,1,\ldots$.
An \agent at site $i$ can perform {\em queries} of the form $q=q(i,j)$
with the following semantics: if site $j$
contains an \agent then the associated answer from site $j$ to the
agent at $i$ is 1 (yes) otherwise 0 (no), $0 \leq i \neq j \leq n-1$. 
For definitional reasons we also require an {\em empty} query $\perp$
(skipped query) with an empty associated answer (skipped answer).
The query and answer take place
in the same time slot.
Given the number $n$ of sites and the number
$k$ of \agents, a {\em mutual search protocol} $A$ 
consists of a (possibly randomized) algorithm to produce 
the sequence of queries
an agent at site $i$ ($0 \leq i \leq n-1$) 
executes together with the time instants it
executes them:
$ A(i) = q_1 ,  \ldots , q_{m_i}$ 
where $q_t$ is the query executed at the $t$th time slot, 
$t := 0, \ldots , m_i$.
If $q_t = \perp$ then
at the $t$th time slot an agent at site $i$ skips a query.
A mutual search {\em execution} of $k$ agents located
at sites $i_1, \ldots , i_k$
consists of the list ${\cal A} = A(i_1), \ldots , A(i_k)$.
We require that in every time slot $t$ there are zero or one
queries from this list that are $\neq \perp$.
Hence we can view ${\cal A}$ as a total order on all queries
by the $k$ agents and $A(i)$ as a restriction to the entries performed by
an agent at site $i$ ($i := i_1, \ldots , i_k$).
The {\em cost} of an execution is the number
of queries $q_t \neq \perp$  with $t \leq t_0$ and $t_0$ is
the least index such that the answer to query $q_{t_0}$ equals 1.
That is, we are interested in the number of queries until first contact
is made. 
\footnote{This is the cost of a {\em non-oblivious} execution.
The cost of an {\em oblivious} execution is simply the number
queries $\neq \perp$ occurring in the list ${\cal A}$.
This case was already completely analyzed in the Introduction.
Therefore, in the remainder
of the paper we only consider non-oblivious executions without
stating this every time.}
The ({\em worst-case}) {\em  cost} of a {\em mutual search protocol} is the
maximum cost of an execution of the protocol.
The {\em worst-case cost} of {\em mutual search} is the
minimum (worst-case) cost taken over all mutual search protocols.

In this paper we consider the case of $k=2$ agents unless explicitly stated
otherwise. The case $k>2$ is open except for the result in 
Section~\ref{multiplayer}.
Informally,
a mutual search protocol specifies, for every site that an \agent
can find itself in, what to do at every time slot: either stay idle
or query another site as specified by the protocol and receive the reply. 
Every time slot harbours at most one query and its answer.
\footnote{We can allow simultaneous queries. If there are $k$
\agents, then every time slot
can have at most $k$ queries $\neq \perp$. The precise cost then
depends only on how we account the at most $k$ queries in the time slot
containing the first query with answer 1. Under different conventions
the results can only vary by $k-1$, that is, by only 1 unit for $k=2$.}

For every pair of sites such a protocol determines
which site will first
query the other. After the first such query takes place,
the  execution terminates, so that the
latter site need never execute the now redundant query of the former site.
Every such algorithm implies a {\em tournament}:
a directed graph having a single arc between every pair
of nodes. An edge from node $i$ to node $j$ represents site
$i$ querying site $j$. The different times at which the  ${n  \choose 2}$
queries/edges are scheduled induce a total timing order on the edges.
Since the cost of running the algorithm depends only on which queries are
made before a certain contacting query, this total order by itself captures
the essence of the timing of queries.

An algorithm can thus be specified by a tournament 
(telling us who queries whom)
plus a separate total timing order on its edges (telling us when). 
Note that the timing order
is completely unrelated to the ordering of the arcs;
the tournament may well be cyclic in the sense of containing
cycles of arcs.
\footnote{For example, algorithm HalfInTurn$_n$ below has cycles
of arcs.}
For us an {\em ordered tournament} is a (tournament, order) pair
where the order is a separate total order on the arcs
of the tournament.

\begin{definition}
An algorithm for MS is an ordered tournament $T=(V,E,\prec)$, where
the set of nodes (sites) is $V=\{0,1,\ldots, n-1\}$, $E$ is a set
of ${n \choose 2} = \half n(n-1)$ edges (queries),
and $\prec$ is a total order on $E$.
For a node $i$, $E_i$ is the set of outgoing edges from $i$,
and is called {\em row} $i$.
The number of queries $|E_i|$ is called the {\em length} of row $i$.
\end{definition}

This way $E_i$ is the set of queries \agent $i$ can potentially make.
Define the cost of an edge as the number of queries that will be
made if the two \agents happen to reside on its incident nodes.

\begin{definition}
The cost $c(T)$ of an algorithm $T$ is the maximum over all edges
$e=(i,j)$ of the edge cost $c(e) = |E_{i}^{\prec e}| + 1 + |E_{j}^{\prec e}|$,
where for any $F \subseteq E$, $F^{\prec e}$ denotes $\{f \in F: f \prec e\}$.
\end{definition}

If the \agents are located at nodes $i$ and $j$ and the edge $e$ between
them is directed from $i$ to $j$, then at the time $i$ queries $j$,
\agent $i$ has made all queries in $E_i^{\prec e}$, while \agent $j$
has made all queries in $E_{j}^{\prec e}$.
We have now all what is needed to present and 
analyze some basic algorithms for the problem
which will form the basis of a better algorithm.

\subsection{Some Simple Mutual Search Algorithms}
\label{simplealgs}

The first algorithm,
AllInTurn$_n$, lets each site in turn query all the other sites.
For instance, AllInTurn$_4$ can be depicted as\footnote{
Another way would be to draw the tournament on nodes $0, \ldots , 3$ and 
label every arc with a time.
The matrix representation
we use seems more convenient to obtain the results.}

\[ \begin{array}{rcccccc}
0: & 1 & 2 & 3 \\
1: &   &   &   & 2 & 3 \\
2: &   &   &   &   &   & 3 \\
3:
\end{array} \]

Here, the sites are shown as labeling the rows of a matrix,
whose columns represent successive time instances.
A number $j$ appearing in row $i$ and column $t$ of the matrix
represents the query from $i$ to $j$ scheduled at time $t$.
As an example execution, suppose the \agents are situated at sites
0 and 2. Then at time 0, (the \agent at site) 0 queries 1 
and receives reply ``no'': the second agent is
not there. Next 0 queries 2 and contacts the second agent, 
finishing the execution
at a cost of 2 queries.

\begin{lemma}
Algorithm AllInTurn$_n$ has cost $n-1$.
\end{lemma}

\begin{proof}
It is in fact easy to see that $c(i,j) = j-i$. A \agent at site $i$ makes
this many queries to contact the other \agent at site $j$, and the latter
never gets to make any queries. The maximum value of $j-i$ is $n-1$.
\end{proof}

A somewhat more balanced algorithm is HalfInTurn$_n$, where
each site in turn queries the next $\lfloor n/2 \rfloor$ sites (modulo $n$).
HalfInTurn$_5$ looks like

\[ \begin{array}{rcccccccccc}
0: & 1 & 2 \\
1: &   &   & 2 & 3 \\
2: &   &   &   &   & 3 & 4 \\
3: &   &   &   &   &   &   & 4 & 0 \\
4: &   &   &   &   &   &   &   &   & 0 & 1
\end{array} \]

For even $n$, sites $n/2 \ldots n-1$ only get to make
$\lfloor n/2 \rfloor -1$ queries.

\begin{lemma}
Algorithm HalfInTurn$_n$ has cost $n-1$.
\end{lemma}

\begin{proof}
Suppose $i<j$. If $j-i \leq \lfloor n/2 \rfloor$,
we find $c(i,j) = j-i$, otherwise $i-j \bmod n = n+i-j$ giving
$c(j,i) = \lfloor n/2 \rfloor + n+i-j$.
Taking $j=i+\lfloor n/2 \rfloor + 1$ achieves the maximum of
$\lfloor n/2 \rfloor + n - (\lfloor n/2 \rfloor+1) = n-1$.
\end{proof}

Our next result shows HalfInTurn$_n$ to be the basis of a much
better algorithm.

\begin{definition}
An algorithm is called {\em saturated} 
if its cost equals its maximum row length.
\end{definition}

An example of a saturated algorithm is AllInTurn$_n$, whose cost of $n-1$
equals the length of row 0.

\begin{lemma}
\label{sat}
An algorithm that is not saturated can be extended with another site
without increasing its cost.
\end{lemma}

\begin{proof}
Let $T$ be an algorithm on $n$ nodes whose cost exceeds all row
lengths. Add a new node $n$, and an edge from every other node to
this new node. Order the new edges after the old edges
(and arbitrarily amongst each other). This does not affect the
cost of the old edges, while the cost of edge $(i,n)$ becomes one more than
the length of row $i$, hence not exceeding the old algorithm cost.
\end{proof}

As the proof shows, the maximum row length increases by exactly one,
so we may add as many sites as the cost exceeds the former.
HalfInTurn$_{2k+1}$ has cost $2k$ and uniform row length $k$ so we
may add $k$ more sites to get a saturated algorithm
SaturatedHalfInTurn$_{3k+1}$ of the same cost:

\begin{corollary}
Algorithm SaturatedHalfInTurn$_n$ has cost $\lceil \frac{2}{3}(n-1) \rceil$.
\end{corollary}

\subsection{Algorithm Refinement}
\label{refinement-sec}

In order to get a better understanding of the structure of 
{\em MS}
algorithms,
we need to focus on their essential properties.
In this section we consider algorithms with only a partial edge ordering.
The question arises how such a partial ordering
can be extended to a good total edge ordering.
The following terminology helps us answer this question.

\begin{definition}
A {\em partial MS}
algorithm is a partially ordered tournament $T=(V,E,\prec,R)$,
where $R \subseteq E$ is the subset of {\em retired} edges,
and $\prec$ is now a partial order, which:
\begin{itemize}
\item totally orders $R$,
\item orders all of $E-R$ before all of $R$, and
\item leaves $E-R$ (pairwise) unordered.
\end{itemize}
An edge $e=(i,j)$ in row {\em prefix} $E_i-R$ has 
{\em retiring} cost $c(e) = |E_i-R| + |E_j-R|$.
Retiring an edge $e$ results in a more {\em refined}
partial algorithm $T=(V,E,\prec',R')$, where $R'=R \cup \{e\}$ and
$\prec' = \prec \cup (E-R',e)$.
\end{definition}

Note that relation $\prec$ is viewed as a set of pairs;
$(E-R',e)$ denotes the set $\{(f,e): f \in E-R'\}$.
The edge $e$ that is added to $R$ was $\prec R$ and since $\prec'$
extends $\prec$, becomes the new earliest edge in $R'$.

Algorithm refinement proceeds backward in time---the queries to be made last
are scheduled first.
An example partial tournament, with 2 retired edges, is
\[ \begin{array}{cccc} (0,3) \\ (0,1) \\ (1,2) \\ (2,0) \end{array} \prec (2,3) \prec (3,1) \]
Note that any sequence of
$|E-R|$ refinements yields a (totally ordered) algorithm,
which we call a total refinement of $T$.
A mere tournament corresponds to a
partial algorithm with no retired edges.

Observe that the cost of $e$ in a total refinement
depends only on its ordering with respect to
the edges in rows $i$ and $j$, which is determined as soon as it retires.
This shows the following

\begin{fact}
\label{retcost}
If $T'$ results from $T$ by retiring edge $e=(i,j)$,
then the retiring cost of $e$ equals the cost of that edge in any
total refinement of $T'$.
\end{fact}

\begin{definition}
The cost $c(T)$ of a partial algorithm $T$ 
is
the minimal cost among all its total refinements.
A total refinement achieving minimum cost is called optimal.
\end{definition}

\begin{lemma}
The cost of a partial algorithm $T$ equals the cost of the partial
tournament that results from retiring the edge $e$
of minimum retiring cost.
\end{lemma}

Informally, any refinement from $T$ will have cost at least
the minimum retiring cost, and choosing $e$ doesn't
hamper us in any way. The following proof makes this notion
of ``non-hampering'' precise.

\begin{proof}
Consider an optimal total refinement from $T$ to some algorithm $T''$,
in which, at some point, say after $e_1, e_2, \ldots, e_k$,
edge $e$ is retired.
Let algorithm $T'$ be the result of retiring $e$ first, and then
continuing the same total refinement with $e$ skipped.
Then $T''$ will have $e \prec'' e_k \prec'' \ldots \prec'' e_1$
whereas $T'$ has $e_k \prec' \ldots \prec' e_1 \prec' e$.
If we compare the costs $c''$ and $c'$ for any edge in
$T''$ and $T'$ respectively, we see that for $1\leq i \leq k$,
$c'(e_i) \leq c''(e_i)$, $c'(e) \geq c''(e)$, and all other edges
cost the same. However, $c'(e) \leq c(e_1)$ by assumption, and
so $T'$ must be optimal too.
\end{proof}

\begin{figure}[p]
\centerline{\epsfxsize=5cm \epsfbox{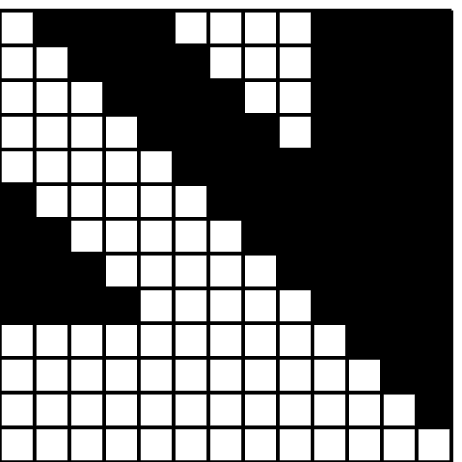}}
\caption{\label{fig13} HalfInTurn$_{13}$}

\end{figure}

Since optimal refinement is straightforward greedy procedure that can
be performed automatically an optimal timed algorithm is uniquely 
determined by just its associated tournament.
By graphically showing the tournament's adjacency matrix,
one obtains a visually insightful representation;
for instance, SaturatedHalfInTurn$_{13}$ is shown in
Figure~\ref{fig13}.

Our algorithm HalfInTurn$_n$ now betrays a bad ordering for even $n$.
It retires $(n-1,0)$ first, at a cost of $n-1$, whereas an optimal
refinement can keep the cost down to $n-2$.
It takes advantage of the bottom rows being shorter,
and first retires an edge between nodes in this bottom half.
For example, the following reordering of HalfInTurn$_4$ has cost 2:
\[ (0,1) \prec (3,0) \prec (0,2) \prec (1,3) \prec (1,2) \prec (2,3) \]

\subsection{Lower Bounds}
\label{lowerbound}

Given that the maximum row length is a lower bound on the cost
of the  algorithm,
the following result is easily obtained.

\begin{lemma}
Every {\em MS} algorithm $T$ for $n$ sites has cost at least
$\lceil \frac{n}{2} \rceil$.
\end{lemma}

\begin{proof}
The average outdegree of a node in $T$ is ${n \choose 2} / n = \frac{n-1}{2}$,
so some row has length at least $\lceil \frac{n-1}{2} \rceil =
\lfloor \frac{n}{2} \rfloor$. It remains to show that
for odd $n$, an algorithm of cost $\frac{n-1}{2}$ is not possible.
This is because for any
collection of $n$ rows each of length $\frac{n-1}{2}$,
the last edge on every row has retiring cost
$\frac{n-1}{2}+\frac{n-1}{2}= n-1$.
\end{proof}

The last argument used in the proof shows that the sum
length of the shortest two rows is a lower bound on an algorithm's cost.
An algorithm of cost $c$ thus necessarily has a row of length at most $c/2$.
Careful analysis allows us to prove the following generalization:

\begin{lemma}
\label{waste}
Let $T$ be an {\em MS} algorithm for $n$ sites with cost $c$.
Then the $(k+1)$st shortest row of $T$ has length at most $c/2 + k$.
\end{lemma}

\begin{proof}
Let $e=(i,j)$ be the last edge for which $i$ and $j$ are not among the
shortest $k$ rows.  Consider the moment of $e$'s retirement in the
refinement from the unordered tournament in $T$ to $T$.  Since $R$
includes at most $k$ edges from each of the rows $i$ and $j$, the
retiring cost of $e$ equals $c(e) = |E_i-R| + |E_j-R| \geq |E_i| - k +
|E_j| - k \geq 2(\min(|E_i|,|E_j|) - k)$. Furthermore, $c(e) \leq c$,
since the cost of $T$ is the maximum of all retirement costs.  It
follows that the smallest of rows $i$ and $j$ has length at most $c/2
+ k$.
\end{proof}

This shows that the best possible distribution of row lengths looks like
{\epsfxsize=3mm \epsfbox{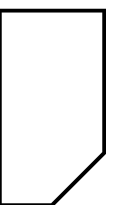}}, where the $n \choose 2$ entries
are divided over $n-c/2$ rows of maximum length $c$, followed by $c/2$
increasingly shorter rows, producing a triangular ``wasted'' space of
size about $(c/2)^2/2$.

\begin{theorem}\label{theo.lbsc}
Every {\em MS} algorithm $T$ for $n$ sites has cost at least
$(4-2\sqrt{3})(n-1)$ {\rm (}$\approx 0.536n${\rm )}.
\end{theorem}

\begin{proof}
Since every row has length at most $c$,
Lemma~\ref{waste} implies:
\begin{eqnarray*}
|E| & = &\frac{n(n-1)}{2} \leq nc - \sum_{k=0}^{c/2} c/2 - k \\
& = & nc - \frac{(c/2)(c/2 + 1)}{2},
\end{eqnarray*}
which in turn implies:
\[ \Rightarrow (c/2)^2 - 2(n-1)c + (n-1)^2 \leq 1 - 1.5c - n \leq 0. \]
Solving for $c$, we find $c \geq (4-2\sqrt{3})(n-1)$.
\end{proof}

\subsection{Algorithm ``Smooth Retiring''}
\label{smooth}

In this section we present our best algorithm, building on the insights
gained in the prevous sections.

Algorithm SR$_n$ is not quite as easy to describe as our earlier algorithms.
It is best described as a partial algorithm with
ordered rows, an optimal refinement of which will be presented
in its cost analysis.

SR$_n$ divides the nodes into two groups: an upper group
$U=\{0,\ldots, u-1\}$ consisting of $u$ nodes
and a lower group $L=\{u,\ldots, n-1\}$ consisting of
$c=n-u$ nodes (which is the cost we are aiming for).
As can be expected, construction of SR$_n$ presumes certain conditions
on the relative sizes of $u$ and $c$, which will be derived shortly.
The value of $c$ will then be chosen as the smallest which satisfies the
conditions.

The upper group engages in HalfInTurn$_u$,
while the lower group engages in a slight variation on AllInTurn$_c$
in which each row is reversed.

Row $u+i$ will have length $c-1-\lfloor \frac{i}{2} \rfloor$, of which
$(u+i,n-1)\ldots (u+i,u+i+1)$ are the last $n-1-(u+i)=c-1-i$ edges.
That leaves $c-1-\lfloor \frac{i}{2} \rfloor-(c-1-i) =
\lceil \frac{i}{2} \rceil$ `slots' available at the front of row $u+i$,
to be filled with edges to $U$.

Row $i<u$ starts with the $\lceil \frac{u}{2} \rceil$ or
$\lfloor \frac{u}{2} \rfloor$ edges in HalfInTurn$_u$, leaving
up to $c-\lfloor \frac{u}{2} \rfloor$ slots per row to be filled with edges
to $L$. The picture so far (with $u=6, c=n-u=8$) is

\[ \begin{array}{rcccccccc}
0: & 1 & 2 & 3 & * & * & * & * & * \\
1: & 2 & 3 & 4 & * & * & * & * & * \\
2: & 3 & 4 & 5 & * & * & * & * & * \\
3: & 4 & 5 & * & * & * & * & * & * \\
4: & 5 & 0 & * & * & * & * & * & * \\
5: & 0 & 1 & * & * & * & * & * & * \\
6: & 13 & 12 & 11 & 10 & 9 & 8 & 7 \\
7: & * & 13 & 12 & 11 & 10 & 9 & 8 \\
8: & * & 13 & 12 & 11 & 10 & 9 \\
9: & * & * & 13 & 12 & 11 & 10 \\
10: & * & * & 13 & 12 & 11 \\
11: & * & * & * & 13 & 12 \\
12: & * & * & * & 13 \\
13: & * & * & * & *
\end{array} \]

Asterisks indicate empty slots. 
By simple geometric properties of the picture we analyze the requirements. 
Define block $B_U$ as the elements in the upper $u$ rows 
determined by $U$ and define block $B_L$ as the elements
in the lower $c=n-u$ rows determined by $L$.
The block $B_U$ has $uc$ elements of which ${u \choose 2}$
are used for the edges in $U \times U$. There
are $uc - {u \choose 2}$ slots in $U$ that can be used for
edges from $U$ to $L$. 
 In the lower block $B_L$ 
the number of open slots that can be used for edges from $L$ and $U$ equals
$(c-1) + (c-3) + \cdots + 2 = (c^2-1)/4$ for odd $c$ and
$(c-1) + (c-3) + \cdots + 1 = c^2/4$ for even $c$.
That is $\lfloor c^2/4 \rfloor$ open slots.
In order to fit all $uc$ edges between $U$ and $L$, the number
of open slots must be sufficient:
\[ uc - {u \choose 2} + \lfloor c^2/4 \rfloor \geq uc .\]
As it happens, the number of elements in $B_U$,
that is $uc$, equals the number of edges between $U$ and $L$.
Therefore,
\begin{equation}
\label{lowerupperfit}
\lfloor \frac{c^2}{4} \rfloor \geq {u \choose 2}
\end{equation}
In the example, the 16 lower slots make up for the
15 which HalfInTurn$_6$ takes out of the top section of size $6 \cdot 8 = 48$.

\subsection{Filling in the slots    }
\label{filling} 

The bottom slots are filled in from top to bottom, left
to right, modulo $u$, starting with $(u+1,0)$. The top slots are then
filled with the remaining edges, in reverse order:

\[ \begin{array}{rcccccccc}
0: & 1 & 2 &  3 & 12 & 10 & 9 & 8 & 6 \\
1: & 2 & 3 &  4 & 12 & 10 & 9 & 7 & 6 \\
2: & 3 & 4 &  5 & 12 & 10 & 8 & 7 & 6 \\
3: & 4 & 5 &  * & 11 & 10 & 8 & 7 & 6 \\
4: & 5 & 0 & 13 & 11 &  9 & 8 & 7 & 6 \\
5: & 0 & 1 & 13 & 11 &  9 & 8 & 7 & 6 \\
6: & 13 & 12 & 11 & 10 & 9 & 8 & 7 \\
7: & 0 & 13 & 12 & 11 & 10 & 9 & 8 \\
8: & 1 & 13 & 12 & 11 & 10 & 9 \\
9: & 2 & 3 & 13 & 12 & 11 & 10 \\
10: & 4 & 5 & 13 & 12 & 11 \\
11: & 0 & 1 & 2 & 13 & 12 \\
12: & 3 & 4 & 5 & 13 \\
13: & 0 & 1 & 2 & 3
\end{array} \]

We assume that $u$ is at least the maximum number of slots per row
$\lfloor c/2 \rfloor$, to avoid filling a row twice with
the same edge:
\begin{equation}
\label{lowerupperfit2}
\lfloor \frac{c}{2} \rfloor \leq u
\end{equation}
This condition also finds use in the next subsection to show optimality
of a certain refinement.

\begin{figure}[p]

\centerline{\epsfxsize=5cm \epsfbox{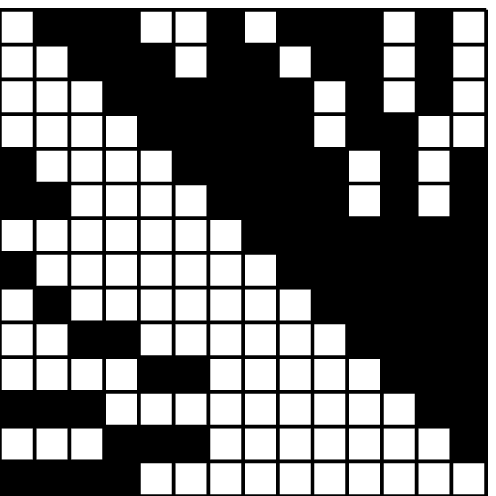}}
\caption{\label{fig14} SR$_{6+8}$}

\end{figure}

\begin{figure}[p]

\centerline{\epsfxsize=6cm \epsfbox{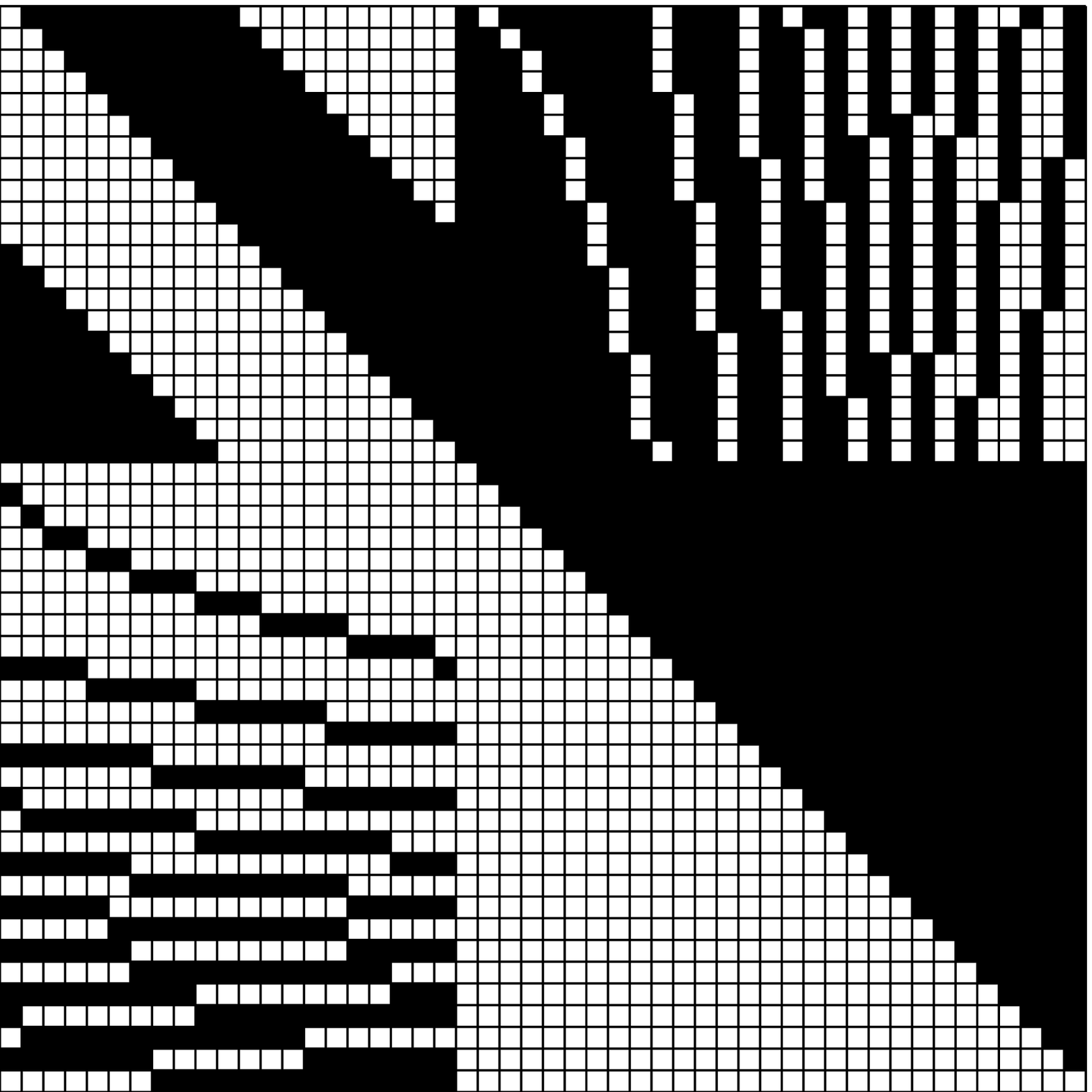}}
\caption{\label{fig50} SR$_{21+29}$}

\end{figure}

The tournament underlying this partial algorithm is shown in
Figure~\ref{fig14}.  Figure~\ref{fig50} makes the pattern clearer with
the bigger instance $u=21, c=29$.

\subsection{Cost analysis}

\begin{theorem}
\label{srcost}
Partial algorithm SR$_n$ has cost $c = \lceil (2-\sqrt{2})(n-1)
\rceil$ {\rm (} $\approx 0.586n${\rm )}.
\end{theorem}

\begin{proof}
To satisfy condition~(\ref{lowerupperfit}), it suffices to have
\[ \frac{c^2}{4} \geq \frac{(n-1-c)^2}{2}, \]
 or equivalently,
$c^2-4(n-1)+2(n-1)^2 \leq 0$, which, solving for $c$,
translates to $c \geq (2-\sqrt{2})(n-1)$.
It remains to show that SR$_n$ actually has cost $c$.
This we do by presenting a total refinement sequence and verifying
all retiring costs.

First, all edges in $L \times L$ are retired, bottom-up and right to left.
Upon retirement of edge $(u+i,u+j)$, $|E_{u+i}-R|$ equals
$|E_{u+i} \cap L \times U| + n-(u+j)$,
while $|E_{u+j}-R|$ equals $|E_{u+j} \cap L \times U|$,
giving a retiring cost of
\[ \lceil \frac{i}{2} \rceil + c-j + \lceil \frac{j}{2} \rceil =
c+ \lceil \frac{i}{2} \rceil - \lfloor \frac{j}{2} \rfloor \leq c,\]
since $i<j$.

Next, all edges $(i,u+j) \in U \times L$ are retired,
in increasing order of $j$.
Upon retirement of edge $(i,u+j)$,
\[|E_i - R| = c- |\{k<j: (u+k,i) \not\in E_{u+k} \}|\]
\[ = c-(j-|\{k<j: (u+k,i) \in E_{u+k} \}|) \leq
c-(j- \lceil \frac{j^2}{4u} \rceil),\]
since the number of slots in the first $j$ bottom rows equals
$(j-1) + (j-3) + \cdots = \lfloor j^2/ 4 \rfloor$, while $i$ appears
once in every $u$ consecutive slots.
Condition~(\ref{lowerupperfit2}) implies
\[  \frac{j}{2u} \leq 
 \frac{c-1}{2u} \leq 1, 
\]
and hence
\[|E_i - R| \leq
c-(j-\lfloor \frac{j}{2} \rfloor \cdot \lceil \frac{j}{2u} \rceil) \leq
c-(j-\lfloor \frac{j}{2} \rfloor) \leq c-\lceil \frac{j}{2} \rceil.\]
Combined with $|E_{u+j} - R| \leq \lceil \frac{j}{2} \rceil$
we conclude $c(i,u+j) = |E_i - R| + |E_{u+j} - R| \leq c$.

Next, all edges in HalfInTurn$_u$ are retired in their usual order
at maximum cost $u-1$, which, by condition~(\ref{lowerupperfit}), is bounded
by $c$.

Finally, all edges in $L \times U$ are retired in arbitrary order,
at costs no more than $\lfloor \frac{c}{2} \rfloor$.
\end{proof}

\section{Asynchronous Case for Two Agents}
\label{asynch}
In an asynchronous setting, one cannot rely on 
queries from different \agents to be coordinated in time.
In some cases the \agents will have no access to a clock, in other cases the
clocks may be subject to random fluctuations. In the asynchronous model,
all an \agent can control, is what other sites are queried, and in what order.
We formalize an asynchronous mutual search (AMS) algorithm as a
partially ordered tournament in which the rows are totally ordered and edges
from different rows are unordered. \footnote{\label{foot.bidirection}
There is a subtlety here. In the synchronous case, we allow only one of
any two given sites to query the other (unidirectional),
reasoning that if both try to query the other,
then one of those queries will always be made first. In the asynchronous case
however, there is no control over which query occurs first, and thus we need
to allow for more general, {\em bidirectional} algorithms (which we refrain
from defining formally here).
Although there may be possible benefits to having two sites query each other,
we have been unable to find ways of exploiting this. We conjecture that
for any bidirectional algorithm, there exists a unidirectional
algorithm of the same or less cost.
Since bidirectional algorithms don't fit too well 
in the existing model, and since
we lack nontrivial results regarding them, 
we use the above unidirectional
definition of AMS algorithm in the remainder of this section.}
The cost of an edge is defined as its position in the row-ordering
(querier cost) plus the length of the target row (queree cost),
since it may happen that the queree has already made all of its queries.

{\bf Upper bound:}
With relatively little control over the ordering of queries, it seems
even less likely to find algorithms which improve on the intuitive bound
of $n-1$ queries. For instance, Lemma~\ref{sat} no 
longer holds in the asynchronous case.
But, surprisingly, a variation of SR$_n$, called ASR$_n$, 
achieves about 1.5 times its cost.
It is obtained by reversing within every row the order of edges
pointing to nodes in
the lower group $L$ of section~\ref{filling}.
The example there
now becomes:

\[ \begin{array}{rcccccccc}
0: & 1 & 2 &  3 & 6 & 8 & 9 & 10 & 12 \\
1: & 2 & 3 &  4 & 6 & 7 & 9 & 10 & 12 \\
2: & 3 & 4 &  5 & 6 & 7 & 8 & 10 & 12 \\
3: & 4 & 5 &  * & 6 & 7 & 8 & 10 & 11 \\
4: & 5 & 0 &  6 & 7 & 8 & 9 & 11 & 13\\
5: & 0 & 1 &  6 & 7 & 8 & 9 & 11 & 13 \\
6: & 7 & 8 &  9 & 10 & 11 & 12 & 13 \\
7: & 0 & 8 &  9 & 10 & 11 & 12 & 13 \\
8: & 1 & 9 & 10 & 11 & 12 & 13 \\
9: & 2 & 3 & 10 & 11 & 12 & 13 \\
10: & 4 & 5 & 11 & 12 & 13 \\
11: & 0 & 1 & 2 & 12 & 13 \\
12: & 3 & 4 & 5 & 13 \\
13: & 0 & 1 & 2 & 3
\end{array} \]

The key observation is that the shortest row has half the length of the maximum
row and that edges to nodes with shorter rows appear in the later positions.
Using an analysis similar to that of Theorem~\ref{srcost}, one arrives at:

\begin{theorem}
\label{asyncsrcost}
Asynchronous algorithm ASR$_n$ has cost at most
$((5-\sqrt{2})/4)n$ {\rm (}$\approx 0.896n${\rm )}.
\end{theorem}

\begin{proof}
We check that every one of the four types of edges
has an asynchronous cost of at most  (note that $u=n-c$):
\[ c+\frac{3}{4}u = \frac{3n+c}{4}, \]
where $c=\lceil (2-\sqrt{2})(n-1) \rceil$ as in Theorem~\ref{srcost}.
For some edges we use the fact that $c \leq \frac{3}{2}u$, and show that
the cost is at most $c+\frac{1}{2}c$.
Recall that row $u+j$ has length $c-1 - \lfloor \frac{j}{2} \rfloor$.
 
Edge $(u+i,u+j) \in L \times L$ has asynchronous cost:
\begin{eqnarray*}
 && \lceil \frac{i}{2} \rceil + j-i-1 + c-1 - \lfloor \frac{j}{2} \rfloor \\
&& = c -2 + \lceil \frac{j}{2} \rceil - \lfloor \frac{i}{2} \rfloor \\
&& \leq c-2 + \lceil \frac{c-1}{2} \rceil.
\end{eqnarray*}

Edge $(i,u+j) \in U \times L$ has asynchronous cost:
\begin{eqnarray*}
&&  \leq \frac{u}{2} + j - |\{k<j: (u+k,i) \in E_{u+k} \}|
+ c-1-\lfloor \frac{j}{2} \rfloor \\
&& \leq c-1+\lceil \frac{j}{2} \rceil + \frac{u}{2} -
\lfloor \frac{j^2}{4u} \rfloor \\
&& \leq c+ \frac{j+u-j^2/2u}{2}. 
\end{eqnarray*}
Writing $j$ as $xu$ gives
\[ j+u-j^2/2u = (x+1-x^2/2)u \leq \frac{3}{2}u, \]
 since $x+1-x^2/2$ assumes
its maximum at $x=1$. Hence, 
\[ c+ \frac{j+u-j^2/2u}{2} \leq c+\frac{3}{4}u. \]
 
Edges  in HalfInTurn$_u$  ($U \times U$) have asynchronous cost at most:
 \[  \frac{u}{2} + c
\leq \frac{3}{2}c.\]
 
Finally, edges in $L \times U$ have cost at most:
\[ \lceil \frac{c-1}{2} \rceil
+ c \leq \frac{3}{2}c.\]
\end{proof}

{\bf Lower bound:} The  $(4-2\sqrt{3})(n-1)$ lower bound on 
the synchronous case (Theorem~\ref{theo.lbsc} holds a fortiori 
for the asynchronous case.

\section{Randomized Case for Two Agents}
\label{random}

For a randomized MS protocol the {\em worst-case expected cost} 
is the worst case, 
over all \agent locations, of the expected
(over the random coin flips) number of queries.
We can use randomization to obtain an algorithm for
mutual search with expected complexity below the proven
lower bound for deterministic algorithms, namely,
a cost of $n/2$.

{\bf Upper bound:}
Algorithm RandomHalfInConcert$_n$ uses the same tournament as
HalfInTurn$_n$, but each \agent randomizes the order of its queries,
and the querying proceeds ``in concert,'' in rounds
that give every row one turn for their next query.
An example where the random choices
have already been made can be depicted as

\[ \begin{array}{rcccccccccc}
0: & 2 &   &   &   &   & 1 &   &   &   &  \\
1: &   & 2 &   &   &   &   & 3 &   &   &  \\
2: &   &   & 3 &   &   &   &   & 4 &   &  \\
3: &   &   &   & 0 &   &   &   &   & 4 &  \\
4: &   &   &   &   & 1 &   &   &   &   & 0
\end{array} \]

\begin{theorem}
Algorithm RandomHalfInConcert$_n$ has a worst-case expected 
cost $ \frac{n+1}{2}$ for $n$ is odd and about $\lceil \frac{n+1}{2} \rceil$
for $n$ is even.
\end{theorem}

\begin{proof}
A worst case occurs when an \agent located at node $n-1$ ends up querying the
other \agent at node $0$ (with the latter already having made a query
in that round). 
The expected number of queries is twice the
number of queries the \agent at $n-1$ makes in a uniformly random order 
of the sites $0,1,\ldots,\lceil \frac{n-1}{2} \rceil$
ending in, and including, the final successful query to site $0$.
This is about  $\lceil \frac{n+1}{2} \rceil$
for $n$ is even and $\frac{n+1}{2}$ for $n$ is odd.
\end{proof}

{\bf Asynchronous Randomized Case:}
Allowing randomness in the algorithm, a $\frac{3n}{4}$ upper bound
is obtained by a variation on RandomHalfInConcert$_n$ in which each row is
ordered randomly. This appears (but is not proven) to be the best one can do.
The best lower bound we have is the synchronous randomized $\frac{n-1}{8}$
lower bound below.

{\bf Lower bound:} 
We prove a lower bound for the synchronous case 
(and hence for the asynchronous case).
\begin{theorem}
For every randomized MS algorithm for two agents on $n$ sites
using a bounded number of coin flips
the worst-case expected cost is at least $\frac{n-1}{8}$.
\end{theorem}
\begin{proof}
Fix a randomized MS algorithm for $n$ sites with $k=2$ agents.
We assume that every agent uses at most a number of coin flips
that is bounded by a total function of $n$. 
Assume that the maximum of the expected number of queries
is $c$ where the  maximum is taken over all placements of two agents
on $n$ sites and the expectation is taken over the randomized coin
flips.
By Markov's inequality, for every placement of the two
agents there is probability at least $\frac{1}{2}$ that
there are at most $2c$ queries by both parties together up to contact.
Hence, at least $\frac{1}{2}$ of all combinations of agents' positions and
sequences of used coin flips have cost at most $2c$.

Now consider a matrix where the rows correspond to the agents'
positions and the columns to the pairs of sequences of coin flips.
It is important for the remainder of the proof 
that two agents at $i,j$ can use sequences of coin flips $\alpha$ and
$\beta$ in two ways: the agent at $i$ uses $\alpha$ and the agent
at $j$ uses $\beta$ and vice versa.
\footnote{Otherwise it can happen that in the same column in agent
positions $(i,j)$ the agent at $i$ uses sequence $\alpha$ and
in the agent positions $(i,k)$ with $k \neq j$ the agent at
$i$ uses sequence $\beta$ ($\beta \neq \alpha$). This would
contradict our intended reduction to a deterministic algorithm.
}
Therefore, there are $2{n \choose 2}$ rows and (by the boundedness
of the length of the coin flip sequence) a bounded number of columns.
By the pigeon hole principle at least one out of all 
coin flip sequence pairs (the columns)
has at least $\frac{1}{2}$ of the $2{n \choose 2}$ agents'
positions (the row entries) incurring cost at most $2c$. 

Fix any such column, say the one determined by coin flip sequences
$\alpha$ and $\beta$. 
We define a ``deterministic pseudo-MS'' algorithm for two agents
in $2n$ sites by splitting every original site $i$
into two copies: a site $i_{\alpha}$ and a site 
$i_{\beta}$ ($0 \leq i \leq n-1$).
Every new site $i_{\gamma}$ ($\gamma \in \{\alpha, \beta \}$) 
executes a new deterministic
algorithm based on the coin flip sequence $\gamma$.
This new algorithm is completely specified by the following:
If the original algorithm 
specifies that an agent at site $i$ using coin flip sequence $\alpha$ 
queries site $j$ 
in time slot $t$, 
then the new algorithm specifies
\footnote{In the actual random execution the pair of coin 
flip sequences in use in this fixed column is $\alpha$ and $\beta$.
Therefore, if there are agents at
site $i$ and site $j$ and the agent at site $i$ uses coin flip
sequence $\alpha$ then the agent at site $j$ uses coin flip sequence
$\beta$.} 
 that an agent at site $i_{\alpha}$
queries site $j_\beta$ in time slot $t$ 
($0 \leq i \neq j \leq n-1$, $t := 1,2 \ldots$). The analogous
rule holds with $\alpha$
and $\beta$ interchanged.
\footnote{Note that $\alpha$-sites only query 
$\beta$-sites and the other way around.}

Consider only the (at least)
${n \choose 2}$ row entries (pairs of nodes) that have cost at most $2c$
and put an arc from node $i_{\alpha}$ 
to node $j_{\beta}$ if $i_{\alpha}$ queries
$j_{\beta}$ in the new algorithm.
This results in a directed bipartite graph on $2n$ nodes
with at least ${n \choose 2}$ arcs.
The average number of outgoing arcs
per node is at least $\frac{n-1}{4}$.  Fix a node with at least
$\frac{n-1}{4}$ outgoing arcs, say $i_{\alpha}$, 
and consider the last node queried by $i_{\alpha}$, say node $j_{\beta}$.
Then $i_{\alpha}$ queries $j_{\beta}$ but $j_{\beta}$ 
doesn't query $i_{\alpha}$.
Hence for the pair $(i_{\alpha},j_{\beta})$ there are at least $\frac{n-1}{4}$
queries executed which, by assumption, is at most $2c$.
Therefore the expectation $c \geq \frac{n-1}{8}$.
\end{proof}

\section{Synchronous Case for Many Agents}
\label{multiplayer}
In the case of $k>2$ \agents we define the mutual search as before,
but now the two \agents involved in a query with an affermative answer, as well as their nodes,
``merge'' into one, sharing all the knowledge they
acquired previously. A query of some node
then becomes a query to the equivalance class of that node.
In this view the goal of the problem is to merge all \agents into one.
\footnote{
Of course, there are other possibilities
to generalize the Mutual Search problem 
$k>2$ \agents, in terms of how \agents that have contacted 
one another coordinate the remainder
of their mutual search.} 

In the two \agent case, an \agent has no input or ``knowledge''
other than the index of the site
he is located at. In the multi-agent case we assume 
a ``full information protocol'' where every \agent is an equivalence class
whose knowledge comprises the complete timed querying and answering 
history of its constituent
\agents. Consequently, algorithms in the
new setting have a vast scope for letting the querying behavior depend
on all those details in case $k>2$. 
Limiting the number of \agents in the new
setting to two reduces exactly to our old model.
\footnote{
We refrain from giving complicated formal
definitions of a multi-player {\em MS} protocol and cost measure
which are not needed for the simple upper bound derived here.}

\ignore{
We now describe algorithm RingSegments.
Algorithm RS$_{n,k}$ splits the $n$-node
search space into a `ring' $R$ of $km(k-1)$ nodes and a
`left-over' group $L$ of $m$ nodes. For simplicity of description
we assume that $n$ is of the form $(k-1)(km+1)$.
}

We now describe algorithm RS$_{n,k}$ (for ``RingSegments'') for $k$ \agents.
The algorithm has a cost below $n$ for all $k=o(\sqrt{n})$. 
Algorithm RS$_{n,k}$ splits the $n$-node
search space into a ``ring'' $R$ of $k(k-1)m$ nodes and a
``left-over'' group $L$ of $m$ nodes. For simplicity of description
we assume that $n$ is of the form $(k(k-1)+1)m$.

The algorithm consists of two phases. During the first phase,
\agents residing on the ring engage in a sort of 
HalfInTurn making $(k-1)m$ queries ahead in the ring.
During the second phase, if not all the \agents are completely joined yet,
\agents query all the leftover nodes.
If, in the first phase, one \agent queries a node affirmatively,
then the \agents merge and the merged \agent continues 
where the front \agent left off,
adding up the number of remaining ring queries of both.
The latter ensures that a collection of $k'$ \agents on the ring
ends up querying $k'(k-1)m$ of ring nodes, with no node queried twice.

\ignore{
For example, consider the case $n=11$ and $b=3$, which yields $k=3$,
a ring of size 9 (nodes $\{0, \cdots ,8\}$), and a segment of 2 
(nodes 9 and 10).
The (initial) picture of the algorithm looks as follows.

\[ \begin{array}{rccccc}
0: & 1 & 2 & 3 & 9 & 10 \\
1: & 2 & 3 & 4 & 9 & 10 \\
2: & 3 & 4 & 5 & 9 & 10 \\
3: & 4 & 5 & 6 & 9 & 10 \\
4: & 5 & 6 & 7 & 9 & 10 \\
5: & 6 & 7 & 8 & 9 & 10 \\
6: & 7 & 8 & 0 & 9 & 10 \\
7: & 8 & 0 & 1 & 9 & 10 \\
8: & 0 & 1 & 2 & 9 & 10 \\
9: & 10 \\
10: & 9
\end{array} \]

\noindent As mentioned above, queries to nodes 9 and 10 are only made
when when the first phase is complete. We now evaluate the cost of the
algorithm.
} 

\ignore{
\begin{lemma} 
Algorithm RS\ has cost $bk$.
\end{lemma}
\begin{proof}
Let $a$ be the number of actual \agents residing on the ring.
Consider first the case $a < b$. 
Then $c(RS) = a (k + \frac{k}{b-1})
\leq (b-1)k + k = bk$. 
Otherwise ($a = b$), the \agents find each other around the ring,
making a total of $bk$ queries in the worst case.
\end{proof}
} 

\begin{theorem} 
Algorithm {\em RS}$_{n,k}$ has cost $k(k-1)m$.
\end{theorem}
\begin{proof}
Let $k'$ be the number of actual \agents residing on the ring.
Consider first the case $k' < k$. Then 
$$
c(\mbox{RS}_{n,k}) = \underbrace{ k'(k-1)m}_{\rm ring \; queries} +
\underbrace{k'm}_{\rm left-over \; queries}$$ $$\leq (k-1)~[(k-1)m + m]
= (k-1)km~.$$ 

Otherwise ($k' = k$), the \agents find each other around the ring,
making
$(k-1)m$ queries each in the worst case.
\end{proof}

\section{Conclusion}
The lower and upper bounds
for the synchronous deterministic two \agent case leave a small gap.
We suspect Lemma~\ref{waste} of being unnecessarily weak.
It is tempting to try and prove a strengthened
version claiming a length of no more than $(c+k)/2$ for the $(k+1)$th
shortest row, which would immediately imply the optimality of SR$_n$.
All algorithms we have looked at so far satisfy this condition.
Unfortunately, there exist simple counterexamples, as witnessed
by row distribution {\epsfxsize=3mm \epsfbox{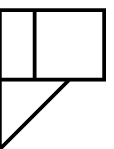}}---where the upper
half engages in a HalfInTurn algorithm before querying the lower half,
which in turn engages in an AllInTurn algorithm (giving a saturated result).
Such algorithms however have lots of relatively short rows, making
them far from optimal. It seems reasonable to expect that an optimal algorithm
has only a constant number of rows shorter than half the cost.
In this light we pose the following conjecture as a lead on optimality of
SR$_n$:
``Let $T$ be an algorithm for $n$ sites with cost $c$,
such that no row is shorter than $\lfloor \frac{c}{2} \rfloor$.
Then the $(k+1)$st shortest row of $T$ has length at most $(c+k)/2$.''

The randomized and asynchronous two-agent cases leave large gaps
between lower bound and upper bound. The multi-agent case
is almost completely unexplored for all models.
The same holds for bidirectional asynchronous algorithms as
in footnote~\ref{foot.bidirection}.

\section*{Acknowledgment}
[HB] dedicates this paper to Kees Buhrman.


\appendix
\section{Related Work}
\label{app.related}

\heading{Distributed Match-Making.}
In ``distributed match-making''~\cite{MV}
the set-up is similar to mutual search
except that if an \agent at node $i$ queries a node $k$ about an agent
residing at node $j$ and the latter agent has posted its whereabout 
at node $k$,
then the query to node $k$ returns $j$~\cite{MV,KV}. In general it is assumed
that the search is in a structured database in the sense that
there have been an initial set of queries from
\agents at all nodes to leave traces of their whereabouts
at other nodes. This problem is basic to distributed mutual
exclusion \cite{Ma} and distributed name server~\cite{MV}.
The difference is that distributed match-making operates in
a cooperative structured environment while mutual search operates in
a noncooperative unstructured environment. Some of our protocol
representation ideas were inspired by this seminal paper.

\heading{Tracking of Mobile Users.} 
Another related search problem is the (on-line) tracking of a mobile user
defined by Awerbuch and Peleg~\cite{AP1,AP2}, where the goal is to access
an object which can change location in the network.
The mobile user moves among the nodes of the network.
\ignore{
~\footnote
{Not exactly
accurate in the case of wire-less, two-tier communication
networks considered in 
subsequent work; we omit further references for succintness.}
} 
From time to time two types of requests are invoked at the nodes:
{\em move}$(i,j)$ (move the user from node $i$ to node $j$) and
{\em find}$(i)$ (do a query from node $i$ to the current location
of the user). The overall goal is to minimize the communication cost.
In contrast, our search problem is symmetric, and the \agents
are static.

\heading{Distributed Tree Construction.}
The goal of {\em MS} can be thought of as forming a clique among
the nodes at which the \agents are located. 
In this sense the problem is related to
tree construction problems, such as the (distributed) 
minimum-weight spanning tree
(MST)~\cite{GHS} and Steiner tree ~\cite{HAKI}. 
Besides other differences 
{\em MS} is concerned
with optimizing the process, and not the outcome of the construction. 

\ignore{
The goal of {\em MS} can be thought of as forming a clique among
the nodes of the \agents. In this sense the problem is related to
tree construction problems, such as the minimum-weight spanning tree
(MST)~\cite{GHS}  and Steiner tree. A main difference is 
that in those problems the nodes are given, and it is the (weight of
the) edges which are (globally) unknown. Additionally, in the case of MST the
tree is supposed to span all the nodes. In the case of
Steiner trees it is enough that the pairs of nodes to be
connected reside in the
same connected component; in the version of {\em MS} we consider 
it is required that the \agents be directly connected.
} 

\heading{Conspiracy Start-Up.}
Another possible application of {\em MS} 
is to secure multi-party computation.
Fault-tolerant distributed computing and secure multi-party
computation
are concerned with $n$ \agents, a fraction ($t$) of which
may be faulty.
It is traditionally assumed ~\cite{BGW,LSP}
that every faulty \agent has complete knowledge of who and where all
faulty agents are, and that they can collude and act in concert. We would like
to weaken this assumption and investigate
the complexity and cost of
achieving such a perfect coordination. We consider
this paper as a first step towards the study of such {\em spontaneous}
adversaries and coalition forming. In fact,
many test-bed problems (Byzantine agreement~\cite{LSP}) and
secure multi-party primitives 
(verifiable secret sharing~\cite{CGMA})
are bound to have interesting characterizations and 
efficient 
solutions
under this new adversary.

\heading{Probabilistic Coalition Formation.}  Billard and
Pasquale~\cite{BP} study the effect of communication environments on
the level of knowledge concerning group, or coalition, formation in a
distributed system. The motivation is the potential for improved
performance of a group of agents depending on their ability to utilize
shared resources. In this particular model the agents make randomized
decisions regarding with whom to coordinate, and the payoffs are
evaluated in different basic structures and amounts of communication
(broadcast, master-slave, etc.).  Their work has in turn been
influenced by work on computational ecologies~\cite{HH} and game
theory studies~\cite{MS}.  In contrast, ours is a search problem with
the goal of minimizing the communication cost of achieving a perfect
coalition.

\heading{Search Theory.}  Finally, {\em MS} is also related
to search theory and optimal search ~\cite{Ko}.  Search theory
is generally concerned with locating an object in a set of $n$
locations, given a ``target distribution,'' which describes the
probability of the object being at the different locations. In turn,
optimal search involves computing how resources (like search time)
can be allocated so as to maximize the probability of
detection. Typically, it is assumed that the target distribution is
known, although more recently this assumption has been
relaxed~\cite{ZO}. Besides the multiple agent aspect, the setting of
{\em MS} is more adversarial, as we measure worst-case cost.


\begin{thebibliography}{99}
\bibitem{AP1} B. Awerbuch and D. Peleg, ``On-line tracking of 
mobile users,'' Technical Memo TM-410, MIT, Lab. for
Computer Science, 1989.

\bibitem{AP2} B. Awerbuch and D. Peleg, ``Sparse Partitions,''
{\em Proc. 31st IEEE Symp. on Foundations of Computer Science},
pp. 503-513, 1990.

\bibitem{AH} R. Axelrod and W. Hamilton, ``The evolution
of cooperation,'' {\em Science}, Vol. 211, pp.~1390-1396, March 1988.

\bibitem{BGW} M. Ben-Or, S. Goldwasser, and A. Wigderson,
``Completeness Theorems for Non-Cryptographic Fault-Tolerant Distributed
Computation,'' {\em Proc. 20th Annual ACM Symp. on the Theory of Computing},
pp. 1-10, 1988.

\bibitem{BP} E. Billard and J. Pasquale,
``Probabilistic Coalition Formation in Distributed Knowledge 
Environments", {\em IEEE
Transactions on Systems, Man, and Cybernetics}, 25(2),  pp.~277-286,
February 1995. 


\bibitem{CGMA}
B. Chor, S. Goldwasser, S, Micali, and B. Awerbuch,
``Verifiable Secret Sharing and Achieving Simultaneity in
the Presence of Faults,''
{\em Proc.  26th Annual IEEE Symposium on the Foundations of
Computer Science}, pp. 383-395, 1985.


\bibitem{GHS}
R. Gallager, P. Humblet and P. Spira,
``A distributed algorithm for minimum-weight spanning trees,''
{\em ACM Transactions on Programming Languages and Systems,}
5(1):66-77, January 1983.

\bibitem{FLP}
M. Fischer, N. Lynch, and M.S. Paterson,
Impossibility of Distributed Consensus with One Faulty Processor.
{\em J. Assoc. Comput. Mach.} 32:2(1985), 374--382.

 
\bibitem{HAKI} S.L.Hakimi, ``Steiner's problem in graphs and its
implications,'' {\em Networks}, 1 (1971) 113-133.

\bibitem{HH} B. Huberman and T. Hogg, ``The behavior
of computational ecologies,'' in {\em The Ecology of
Computation} (B. Huberman, ed.), North Holland, Elsevier Science
Publishers, 1988.

\bibitem{Kn98}
D.E. Knuth, {\em The Art of Computer Programming, Vol. 3:
Sorting and Searching}, Addison-Wesley, 1998.

\bibitem{Ko}
B.O. Koopman, ``The Theory of Search, Parts I--III,''
{\em Operations Research} Vol. 4, pp.~324-346 (1956),
Vol. 4, pp.~503-531 (1956), and Vol. 5, pp.~613-626 (1957).

\bibitem{KV}
E. Kranakis and P.M.B. Vitanyi, 
``A note on weighted distributed Match-Making,''
{\em Mathematical Systems Theory}, 25(1992), pp.~123-140. 

\bibitem{LSP}
L.~Lamport, R.E.~Shostak and M.~Pease,
``The Byzantine generals problem,''
{\em ACM  Trans.\ Prog.\ Lang.\ and Systems}, 4:3 (1982),
pp.~382--401.

\bibitem{Ma}
M. Maekawa, 
``A $\sqrt{N}$ Algorithm for Mutual Exclusion
in Decentralized Systems,'' {\em ACM Transactions on Computer
Systems}, 3(1985), pp. 145-159.

\bibitem{MS}
J. Maynard-Smith,
{\em Evolution and the Theory of Games}, Cambridge University Press,
1982.

\bibitem{MV}
S.J. Mullender and P.M.B. Vitanyi, ``Distributed match-making,''
{\em Algorithmica}, 3 (1988), pp.~367-391.


\bibitem{ZO}
Q. Zhu and J. Oommen, ``Optimal Search with Unknown Target
Distributions,'' to appear in {\em Proc. XVII International Conference
of the Chilean Computer Science Society}, Valpara\'{i}so, Chile,
November 1997.

\end{thebibliography}
\end{document}